\documentclass[fleqn,usenatbib]{mnras}
\usepackage{newtxtext,newtxmath}

\usepackage[T1]{fontenc}

\DeclareRobustCommand{\VAN}[3]{#2}
\let\VANthebibliography\thebibliography
\def\thebibliography{\DeclareRobustCommand{\VAN}[3]{##3}\VANthebibliography}

\usepackage{graphicx}	
\usepackage{tabularx}
\usepackage{amsmath}	
\usepackage{indentfirst}
\usepackage{appendix}
\usepackage{makecell}
\usepackage{enumerate}
\usepackage{enumitem}
\setlist[enumerate]{labelwidth=0pt, labelsep=0.5em}
\usepackage{multirow}
\usepackage{gensymb}
\usepackage{longtable}
\usepackage{pdflscape}
\usepackage{float}
\usepackage{lineno}
\usepackage{scrextend}
\usepackage{threeparttable}
\usepackage{lastpage}

\title[ISM down to the central kpc of quasar hosts at $z=2$]{Under the glare of a luminous quasar, the FIR continuum is still an excellent tracer of the ISM down to the central kiloparsec}

\author[Silverman et al.]{
\parbox{\textwidth}{
\Large
John D. Silverman$^{1, 2, 3, 4}$\thanks{E-mail: john.silverman@ipmu.jp}
Emanuele Daddi$^{5}$
Qing-Hua Tan$^{6}$
Zhaoxuan Liu$^{1, 2}$
Qinyue Fei$^{7, 8}$
Francesco Valentino$^{9, 10}$
Luis C. Ho$^{7, 8}$
Vincenzo Mainieri$^{11}$
Jed McKinney$^{12}$
and Wiphu Rujopakarn$^{13, 14}$
}
\\
\\
\parbox{\textwidth}{$^{1}$Kavli Institute for the Physics and Mathematics of the Universe (Kavli IPMU, WPI), UTIAS, Tokyo Institutes for Advanced Study, University of Tokyo, Chiba, 277-8583, Japan\\
$^{2}$Department of Astronomy, School of Science, The University of Tokyo, 7-3-1 Hongo, Bunkyo, Tokyo 113-0033, Japan\\
$^{3}$Center for Data-Driven Discovery, Kavli IPMU (WPI), UTIAS, The University of Tokyo, Kashiwa, Chiba 277-8583, Japan\\
$^{4}$Center for Astrophysical Sciences, Department of Physics \& Astronomy, Johns Hopkins University, Baltimore, MD 21218, USA\\
$^{5}$Université Paris-Saclay, Université Paris Cité, CEA, CNRS, AIM, F-91191 Gif-sur-Yvette, France\\
$^{6}$Purple Mountain Observatory, Chinese Academy of Sciences, 10 Yuanhua Road, Nanjing 210023, PR China\\
$^{7}$Kavli Institute for Astronomy and Astrophysics, Peking University, Beijing 100871, People's Republic of China\\
$^{8}$Department of Astronomy, School of Physics, Peking University, Beijing 100871, People's Republic of China\\
$^{9}$Cosmic Dawn Center (DAWN), Copenhagen, Denmark\\
$^{10}$DTU Space, Technical University of Denmark, Elektrovej 327, DK2800 Kgs. Lyngby, Denmark\\
$^{11}$European Southern Observatory, Karl-Schwarzschild-Strasse 2, Garching bei M\"{u}nchen, Germany\\
$^{12}$Department of Astronomy, The University of Texas at Austin, 2515 Speedway Boulevard Stop C1400, Austin, TX 78712, USA\\
$^{13}$National Astronomical Research Institute of Thailand, Don Kaeo, Mae Rim, Chiang Mai 50180, Thailand\\
$^{14}$Department of Physics, Faculty of Science, Chulalongkorn University, 254 Phayathai Road, Pathumwan, Bangkok 10330, Thailand\\
}
}

\date{Accepted XXX. Received YYY; in original form ZZZ}

\pubyear{2025}

\begin{document}
\label{firstpage}
\pagerange{\pageref{firstpage}--\pageref{lastpage}}
\maketitle


\begin{abstract}

Contamination-free assessments of the interstellar medium (ISM) and star formation in quasar host galaxies, particularly based on far-infrared (FIR) emission, offer insights into the role of supermassive black holes in galaxy evolution. Motivated by recent predictions from simulations of quasar heating of dust on both nuclear and galaxy-wide (extended) scales, we perform two-component (host galaxy + point source) modeling of high-resolution ($\sim0.1^{\prime\prime}$) ALMA observations of the FIR continuum in Band 5 ($\lambda_{rest}\sim500~\mu m$) of three highly luminous quasars ($L_{bol}\sim10^{47}$ erg s$^{-1}$), powered by supermassive black holes having M$_{BH}\sim10^9$ M$_{\odot}$, at $z=2$. For two quasars, we include Band 9 ($\lambda_{rest}\sim154~\mu m$; 0.06$^{\prime\prime}$ and 0.3$^{\prime\prime}$) data at high signal-to-noise which places further constraints on the unresolved nuclear component in each case. To break the degeneracy between quasar and stellar heating of dust on extended scales, we use the molecular line CO (J=5--4), observed in Band 5, to gauge the expected contribution of star formation to the infrared luminosity. We find very good agreement, between the strength and spatial distribution of the extended continuum component and its prediction based on the CO (J=5-4) emission. This is supported by the location of our three quasars along the $L_{CO (5-4)}-L_{IR, SFR}$ luminosity relation for inactive star-forming galaxies. As a consequence, there is no evidence for additional continuum emission on extended scales which could be attributed to quasar-heated dust. As expected, the nuclear (i.e., torus) contribution is present and subdominant (e.g., 12\% in Band 9 for one quasar with a typical star-forming host) or non-existent (e.g., <8\% in Band 9 for the starbursting host). Based on both the continuum and CO, the presence of substantial levels of ongoing star formation agrees with previous estimates from unresolved ALMA continuum observations which finds SFRs consistent with star-forming main-sequence galaxies. Therefore, our results do not provide evidence for a quasar-mode feedback, even for the most luminous cases at $z=2$.

\end{abstract}


\begin{keywords}
quasars: supermassive black holes -- galaxies: ISM -- galaxies: active -- submillimter: ISM
\end{keywords}






\section{Introduction}

Ever since the discovery of quasars, there has been an intense effort to determine which types of galaxies host supermassive black holes (SMBHs) across a range of mass, accretion rate, and cosmic time. Early studies identified massive elliptical galaxies as the hosts of powerful radio galaxies \citep[e.g.,][]{Disney1995,Dunlop2003}. Since then, studies of optically-selected (i.e., radiatively-efficient) AGN, and the more luminous QSOs, including those at higher redshifts find the hosts to be massive galaxies and forming stars at rates more-or-less consistent with the typical star-forming populations. These results are based on a range of complementary investigations using optical spectroscopic surveys such as SDSS \citep{Kauffmann2003} and zCOSMOS \citep{Silverman2009}, ground-based imaging \citep[e.g.,][]{Li2021,Zhuang2022}, $Hubble$ Space Telescope imaging \citep[e.g.,][]{Jahnke2004,Zhao2021}, $Spitzer$ in the mid-IR \citep[e.g.,][]{Xie2021}, and $Herschel$ in the FIR \citep[e.g.,][]{Mullaney2012,Stanley2017}. In all these studies \citep[see][for a review]{Harrison2024}, there is still concern that the emission has a non-negligible component from the AGN that may not fully be accounted for, thus galaxy properties, particularly star formation rates (SFRs), may be inaccurate.

Over the past decade, it has been typically assumed that FIR wavelengths, accessible with Atacama Large Millimeter Array (ALMA) and the Northern Extended Millimeter Array (NOEMA), enable a quasar-free window on the surrounding gas and dust emission to assess the content, distribution and kinematics of the ISM in quasars hosts out to $z\sim6$. Many studies have inferred the SFRs of quasar hosts through either line (e.g., [CII], higher-order CO transitions) or continuum emission \citep[e.g.,][]{Bertoldi2003,Bischetti2018,Izumi2018,Bischetti2021,Venemans2018,Xie2021, Lamperti2021}.

For example, the dust continuum emission at rest-frame 285$\mu$m was measured for 20 of the most luminous ($log~L_{bol}$ $>$ 46.9; units of erg s$^{-1}$) quasars at $z\sim2$ to determine the total SFRs of their host galaxies \citep{Schulze2019}. These quasars are powered by SMBHs having $log~M_{BH}$ $>$ 9.2 ($M_{\odot}$). It is at these high luminosities where we would expect radiative-mode feedback to impact the ISM of their host galaxies, if not at lower luminosities. The FIR emission, used to measure the total thermal emission, potentially establishes a galaxy-wide SFR based on a $\sim1^{\prime\prime}$ beam (7.5 kpc at $z=2$) and an assumed dust temperature ($T_{dust}=47~K$) which can lead to considerable uncertainty for individual cases. With this sample, the SFR distribution, both the mean and dispersion, is equivalent to that of massive ($\sim10^{11}$ M$_{\odot}$) star-forming galaxies at the same redshift. If true, there is no evidence to support the presence of quasar-driven feedback impacting the gas in a manner to shut down galaxy-wide star formation on timescales greater than 100 Myrs. There could still be effects star formation from AGN outflows on shorter timescales as shown by simulations.    

However, a correction to the continuum emission, attributed to dust heated by the quasar, may not have been accurately taken into consideration when determining the SFR. This is due to the fact that available SEDs \citep{Elvis1994,Lyu2017a,Lyu2017b} for quasars are based on models with assumed temperature distributions that are susceptible to uncertainties in the FIR, particularly at wavelengths ($>100$ $\mu m$), despite the rapid drop of the AGN thermal emission. Any change in slope or normalization of the AGN SED at FIR wavelengths \citep{Perna2018,Kirkpatrick2019} will impact the assessment of the emission attributed to star formation. 

\citet{McKinney2021} investigated the capability for an AGN to influence the cold dust emission using merger simulations with radiative transfer models. They claim that an AGN can heat the cold dust, above that produced by the quasar torus, by factors of 2--3$\times$ in the FIR ($>$ 100$\mu$m). While the enhancement is seen primarily for a nuclear component, there is excess heating on the larger scale of the host galaxy. Further efforts \citep{Schneider2015,DiMascia2021} address similar issues for luminous quasars at $z\sim6$. If this AGN component to the FIR emission is realized for the ALMA studies discussed above \citep{Schulze2019,Scholtz2021}, the SFR distribution could shift below that of the SF MS thus providing evidence for the role of SMBHs in quenching SF in massive galaxies. 

In support of this scenario, \citet{Tsukui2023} analyzed the spatial distribution of dust temperature using two FIR bands in a HyLIRG at $z=4.4$. They find a central component significantly warmer ($57.1\pm0.3$ K) than the larger host galaxy ($\sim38$ K) which is attributed to AGN heating. As a result, the estimate of the SFR is reduced by a factor of 3 while still maintaining its starburst nature. However, further spatially-resolved studies with ALMA are needed to assess the AGN contribution to the total FIR emission and the effectiveness of using ALMA as a 'clean' SFR machine for AGN host galaxy studies, including the less extreme cases in terms of their host properties (i.e., SFR).

To expand on spatially-resolved studies of quasar hosts with ALMA, we present observations of the FIR continuum and CO J=5--4 emission at kiloparsec resolution, and below, of three luminous quasars at $z=2$ from the sample of \citet{Schulze2019}. We assess the nature of the FIR continuum including a central contribution from the quasar using both ALMA bands 5 and 9. The CO J=5--4 emission line, shown to trace dense gas and strongly correlate with the total infrared luminosity in high-redshift star-forming galaxies \citep{Daddi2015}, is used as an independent proxy to assess the level of ongoing star formation, even in the inner kpc. Throughout this work, we use a Hubble constant of $H_0 = 70$ km s$^{-1}$ Mpc$^{-1}$ and cosmological density parameters $\Omega_\mathrm{m} = 0.3$ and $\Omega_\Lambda = 0.7$. We assume a Chabrier initial mass function for estimates of mass and star formation rate.

\section{Method}
\label{text:methods}

Our aim is to confirm (or not) whether FIR emission, measured on galaxy-wide scales, as previously reported in \citet{Schulze2019}, can be primarily attributed to ongoing star formation in the quasar host galaxy. Evidence for a quasar-heated component can be found if the majority of the continuum emission is highly compact or there is a significantly extended component with weak or non-existent CO emission at the same physical scale. If realized, the galaxy-wide SFR estimates will need to be appropriately lowered.

Specifically, we use the extended component in CO (J=5--4) to infer the contribution of star formation to the infrared continuum luminosity thus placing constraints on the dust being heated by the quasar, i.e., breaking the degeneracy between quasar and stellar heating. This is enabled by the tight (within 0.2 dex) relation between CO (J=5--4) and total infrared luminosity \citep{greve_star_2014,Daddi2015,liu_high-j_2015,Valentino2020}, even in AGN host galaxies \citep{Valentino2021}.

For this purpose, ALMA Band 5 ($\nu \sim205$ GHz: observed) observations were carried out in Cycle 9 with an antenna configuration resulting in spatial resolution of $0.1^{\prime\prime}$. For three luminous SDSS quasars at $z\sim2$, we measure the FIR continuum emission ($\lambda_{rest}\sim490$ $\mu$m at $z=2$) and CO (J=5--4; $\nu_{rest}=576.27$ GHz). Band 9 observations (0.06 and 0.3$^{\prime\prime}$) are available for two of the three quasars to further differentiate the extended continuum from the presence of an unresolved nuclear component to the FIR emission as further detailed below.

\subsection{Targets}

For this test, we present three of the most optically-luminous ($log~L_{bol}\sim47$; units of ergs s$^{-1}$) and radio-quiet SDSS quasars at $z=2$ from the sample of  \citet{Schulze2019}. These quasars (SDSS J1236+0500, SDSS J2317-1033 and SDSS J2345-1104; see Table~\ref{tab:sample} for full names) are powered by SMBHs having $M_{BH}\sim2-3\times10^9$ M$_{\odot}$ based on the MgII emission line and single-epoch virial method. We recognize that these may be overestimates given the recent results from GRAVITY \citep{Abuter2024} and discuss the implications which do not impact our analysis of the ISM properties of their hosts. Initial estimates of their SFRs are measured using ALMA continuum observations in Band 7 (Cycle 5; 2017.1.00102.S; PI: A. Schulze) with a spatial resolution of 0.9$^{\prime\prime}$ \citep{Schulze2019}. For this study, three quasars are chosen for analysis at higher spatial resolution with different levels of FIR luminosity in Band 7; two quasars (J2317-1033, J2345-1104) have luminosities at $z=2$ typical of SF MS galaxies ($log~L_{850\mu m}\sim44.1$; units of ergs s$^{-1}$) while J1236+0500 is $5\times$ higher ($log~L_{850\mu m}\sim44.8$; units of ergs s$^{-1}$) thus considered a starburst galaxy. Their basic properties from the Band 7 data are listed in Table~\ref{tab:sample}.

\begin{figure*}
\begin{centering}
\includegraphics[width=0.9\textwidth]{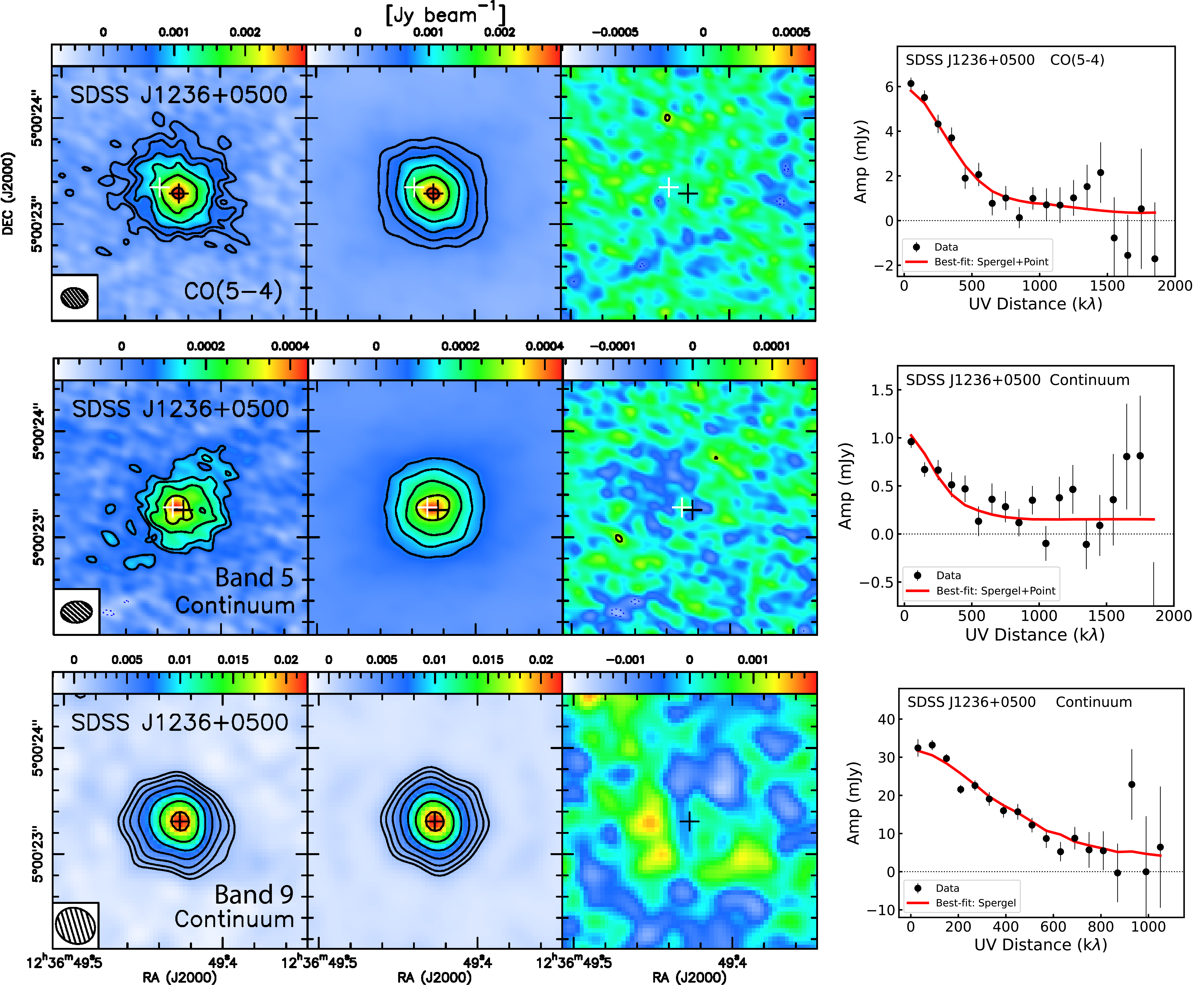}
\end{centering}
\caption{Two-component (Spergel+point source) modeling the ALMA imaging of the star-bursting quasar host of SDSS J1236+0500 for CO (J=5-4) and the continuum (Bands 5 and 9): Best-fit results from model profile fitting in the $uv$-plane to CO(J=5--4) line emission are shown on top while the results for the band 5 and 9 continuum emission are given below. From left to right, we show the dirty image, source model convolved with the dirty beam and, and residuals after subtracting the source model. Contours start at $+3\sigma$ and increase by a factor of 1.5. The 1$\sigma$
rms noise levels for the CO(5-4) and continuum maps are 114 $\mu$Jy beam$^{-1}$, 29 $\mu$Jy beam$^{-1}$ (Band 5) and 0.46 mJy beam$^{-1}$ (Band 9) based on velocity widths reported in Table~\ref{tab:band5_9}. The white and black crosses mark the centers of the point and Spergel components, respectively. See text for details on model parameters and constraints. $Right$ column: Amplitude versus UV distance (data points) for CO (J=5-4) and the continuum with best-fit model shown by the red curve.}
\label{fig:model_fit_J1236}
\end{figure*}

\subsection{Data}

ALMA observations of three quasars at $z=2$ were executed in Cycle 9+10 (Projects 2022.1.00707.S and 2023.1.00185.S; PI Silverman) using 42-44 12m antennas and the Band 5 receiver tuned to frequencies between 190 and 209 GHz with a resolution of 3.9 MHz and spectral bandpass of 1.875 GHz for each of the four spectral windows. Both high ($0.1^{\prime\prime}$; Cycle 9) and low ($1^{\prime\prime}$; Cycle 10) spatial resolution observations were obtained. At high (low) resolution, on-source exposure times were 24.8 (15.1), 102.6 (52.5), and 138.8 (47.4) min for J1236+0500, J2317-1033 and J2345-1104 repectively. Standard targets were used for flux, bandpass and phase calibration. The data were prepared for science with the CASA pipeline versions 6.4.1.12 (J1236+0500) and 6.8.4.9 (J2317-1033, J2345-1104) for the high-resolution observation while CASA version 6.5.4.9 was used for the low resolution. For each quasar, the high and low resolution visibilities are coadded using the CASA task `concat'.

As mentioned, a subset of 12m observations in Band 9 from the Cycle 11 ALMA program (2024.1.00596.S; PI Silverman) were completed. J1236+0511 was observed at a representative frequency of 658.9 GHZ for 5.1 min at a spatial resolution of 0.35$^{\prime\prime}$ (2.9 kpc) while J2345-1104 was observed at 658.35 GHz for 31.7 min at 0.06$^{\prime\prime}$ (0.5 kpc). For both targets, eight spectral windows covered 1.98 GHz each with a spectral resolution of 31.25 MHz. These data have high signal-to-noise enabling two component modeling of the continuum emission which places important constraints on the presence of an unresolved nuclear component. Furthermore, these observations include the [CII] line which will be presented in a future study of the gas kinematics.

\subsection{Model fitting in the $uv$-plane}
\label{sec:method}

We perform model fitting in the $uv$-plane to determine the flux and half-light radii of the CO (J=5--4) and continuum emission thus mitigating inherent systematic biases when fitting sources in the image plane. First the coadded visibilities are exported from CASA and imported to GILDAS \citep{Guilloteau2000} using exportuvfits and fits.to.uvt tasks. Two components are included in the model to account for an extended (disk-like) and a central unresolved (quasar) component. 

The extended source is described by an elliptical Spergel profile \citep{Spergel2010} which is an analytic alternative to a Sersic profile with forms in both real and Fourier space thus able to be implemented in $uv$-space which avoids issues with working with images based on a clean algorithm. The use of this function has been well documented \citep{Tan2024a} and implemented in the study of galaxies at cosmic noon \citep{Tan2024b}.

\begin{figure*}
\begin{centering}
\includegraphics[width=0.9\textwidth]{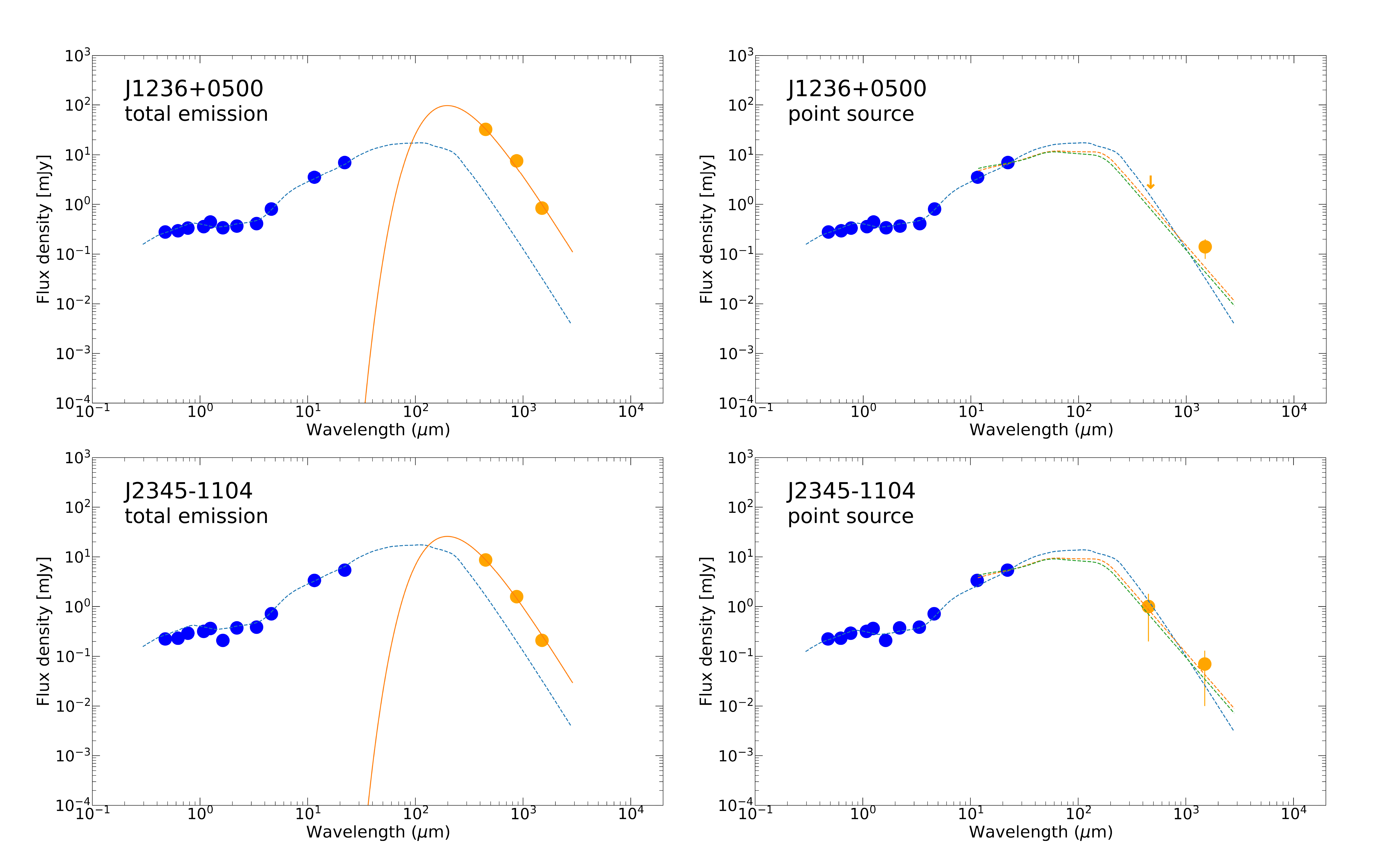} 
\end{centering}
\caption{Multi-band (5, 7, 9) continuum imaging for SDSS J1236+0500 (top row) and SDSSJ 2345-1104 (bottom row): (left) Observed-frame total broad-band SED with photometry from SDSS and WISE in blue with the quasar template from \citet{Lyu2017a} overlaid. ALMA continuum fluxes in orange are in agreement with a modified blackbody spectrum at T=47 K, well above the quasar template. (right) ALMA continuum flux and limit for the unresolved central source in Bands 5 and 9 with $2\sigma$ errors. In addition, warm-dust deficient and hot-dust deficient quasar templates from \citet{Lyu2017a} are overlaid and normalized to match the 22.5$\mu$m WISE photometry. In both quasars, the unresolved continuum is primarily consistent with emission from the torus.}
\label{fig:seds}
\end{figure*}

\section{Results}

Our approach is to first describe the results for J1236+0500 since the ALMA maps are of substantially higher signal-to-noise, for CO (J=5--4) and the FIR continuum, given the starburst nature of its host galaxy which further allows us to model the galaxy kinematics. Then we present the ALMA observations of two quasars (J2317-1033, J2345-1104) with hosts having more typical dust and gas properties, similar to main-sequence (MS) star-forming galaxies at $z=2$. 

For the two categories (starburst, MS), the properties of the CO (J=5--4) emission are presented first for individual sources since the interpretation of the continuum emission relies on the CO as described in Section~\ref{text:methods}. Each tracer (continuum, line) will then be assessed and compared as indicators of the SFR in the host galaxies of luminous quasars after the removal of an unresolved component likely due to further gas excitation and dust heating by the quasar. For two quasars, the FIR imaging in band 9 allows us to better assess the contribution of the central unresolved component which is attributed to the dusty torus. Lastly, the dust and CO emission from these three quasar hosts will be compared with a compilation of star-forming galaxies and lower-luminosity AGNs at $1<z<1.7$ \citet{Valentino2021} to conclude on the quasar contribution to the total FIR emission.

\subsection{SDSS J1236+0500: a luminous quasar in a starbursting host}
\label{text:j1236}

As shown in the top row of Figure~\ref{fig:model_fit_J1236}, CO (J= 5--4) emission is significantly detected with a $S/N=15.8$. The CO image is produced by using Briggs weighting scheme (2.0) and summing the cube over a velocity range ($\Delta$v), given in Table~\ref{tab:band5_9}, which is centered on the CO line and includes all velocity channels with detectable flux. The rms noise level is 0.08 Jy beam$^{-1}$ km s$^{-1}$. In the top-left panel (uncleaned image), the CO emission is spatially-extended with a central peak as shown by the contours starting at 3$\sigma$. The amount of CO emission is indicative of a large reservoir of molecular gas as expected based on the bright FIR continuum previously detected in Band 7 (Table~\ref{tab:sample}). 

We model the source in the $uv$-plane (Section~\ref{sec:method}) using the task $'uv\_fit'$ in GILDAS to quantify the emission strength and its spatial distribution. Two components are included to account for an unresolved central source and an extended galaxy component. A Spergel profile is implemented to fit the extended emission.  For this exercise with J1236+0500, we leave the index of the extended component free which results in $\nu=0.34\pm0.43$ which is equivalent to a S\'ersic profile with $n_s\sim1$ (an exponential disk; see Figure 2 of \citealt{Tan2024a}). While a better constraint on $\nu$ requires higher S/N data \citep{Tan2024a}, there is little change in the profile shape over the range of uncertainty, as shown in the same figure. In the top row, the middle and right panels show the best-fit model and residual image. The latter demonstrates that an extremely good fit can be achieved when modeling the emission in the $uv$-plane with a smooth model which accounts for essentially all of the extended emission seen in the uncleaned image.  In Table~\ref{tab:band5_9}, we provide the measurements based on this model fit including the total CO (J=5-4) luminosity ($L^{\prime}$) of $2.2\pm0.4\times10^{10}$ $L_{\odot}$, which is determined as:
\begin{equation}
L'_{\rm line}\,[\mathrm{K\,km\,s^{-1}\,pc^2}] = 3.25\times10^7\,S_{\rm
  line}\,\Delta v\, \nu_{\rm obs}^{-2}\,(1+z)^{-3}D_{\rm L}^2
,\end{equation}
where $S_{\rm line}\,\Delta v$ is the velocity-integrated line flux in
Jy km s$^{-1}$, $v_{\rm obs}$ is the observed line frequency in GHz, $z$ is
the redshift, and $D_{\rm L}$ is the luminosity distance in Mpc. 

In particular, we find that the majority (94.3\%; $5.82\pm0.26$ mJy) of the CO emission is best fit by an extended component having an effective radius of $1.01\pm0.08$ kpc.
The extended nature is demonstrated by the drop in amplitude at larger uv distances in Figure~\ref{fig:model_fit_J1236}. There is also an unresolved component, offset from the galaxy center by the size of the beam, which is significantly detected at $0.35\pm0.08$ mJy. However, the reality of this component is questionable since a fit with the point-source component, fixed to a common center with the Spergel profile, results in a negative flux. Therefore, we conclude that the unresolved emission is minimal at best in this case and would not dramatically affect lower-resolution observations based on the global CO (J=5-4) emission. We conclude that the significant detection of extended CO (J=5--4) on galactic scales reflects the presence of dense molecular gas which is forming stars at prodigious rates (as further assessed in Section~\ref{text:sfrs}).

\subsubsection{Continuum: an assessment of quasar contribution to the thermal emission}

With CO molecular gas present and likely forming stars, we examine the FIR continuum emission in band 5 and 9 to determine whether it provides a similar assessment on the stellar growth rates and exhibits two distinct model components. In the middle panels of Figure~\ref{fig:model_fit_J1236}, the spatial distribution of the FIR emission observed at $\nu$=204 GHz (Band 5; $\lambda\sim450~\mu$m in the rest-frame) for J1236+0500 is shown as done for CO (J=5--4). These are based on the sum of continuum emission over the full bandwidth (1.8 GHz) of the four spectral windows with the exception of the channels including CO emission. The rms noise level is 0.029 mJy beam$^{-1}$. As shown the FIR emission is significantly detected with at least an extended component thus warranting two-component modeling.

Specifically, we fit the continuum using a Spergel model for the extended emission and a point-source for a potentially unresolved component (Fig.~\ref{fig:model_fit_J1236}). As done for CO, the task $uv\_fit$ is implemented to avoid systematic effects from fitting in the image plane. We fix the Spergel index to the best-fit value ($\nu=0.3$) from the CO fit. The total continuum emission is detected at a high level of significance (S/N of 12.5) and has an integrated flux of $S_{205~GHz}$=1.07$\pm$0.08 mJy (Table~\ref{tab:band5_9}). We find a significant detection ($0.15\pm0.03$ mJy) of an unresolved component which accounts for 14.2\% of the total flux. In Fig.~\ref{fig:model_fit_J1236}, the smooth model nicely accounts for the majority of the emission as shown by the lack of any strong residuals. Therefore, the nuclear component to the FIR emission, including a likely contributions from the torus and central star formation,
is marginal thus not dominating the total emission.

We further demonstrate that an unresolved component is subdominant by displaying the amplitude as a function of uv distance for the data and best-fit model (Fig.~\ref{fig:model_fit_J1236}). At larger uv distances, the amplitude drops and approaches zero relative to the peak on the shortest baselines (i.e., largest scales). This is as expected for a source primarily with extended emission. The data points at the largest UV distance are affected by low S/N.

These results are supported by a Band 9 observation with a beam size of 0.30$^{\prime\prime}$ (bottom panels of Figure~\ref{fig:model_fit_J1236}). From the two-component modeling in $uv$-space as done above, an unresolved central component is not detected (Table~\ref{tab:band5_9}). Here, we fix the Spergel index to 0.3 as determined from the fit to the CO (J=5-4) emission. we are only able to place an upper limit on an unresolved component at 2.8 mJy (3$\sigma$) which represents at most 8\% of the total continuum emission (34.6 mJy). Reassuringly, the extended emission in band 9 has a similar effective radius at 1 kpc and the flux, along with bands 5 and 7, is consistent with a thermal spectrum (i.e., modified black body with $\beta=1.6$ and T=47 K), well above the standard quasar template \citep{Lyu2017b}, as shown in the top-left panel of Figure~\ref{fig:seds}. In the top-right panel, the decomposed flux of the central unresolved component in Band 5 and upper limit in band 9 are nearly consistent with emission expected from the quasar model of \citet{Lyu2017a}.

As shown in Fig.~\ref{fig:seds} (top-right panel), there may be excess unresolved emission in band 5 due to dust heated by the quasar or compact star formation which is not associated with the torus. It is also possible that the torus may lack hot or warm dust as shown by the additional quasar templates which have a shallower slope than the standard quasar and provide a better match to the ALMA constraints at the expense of over predicting the fluxes in the WISE bands. In any case, we conclude that any unresolved continuum component is weak and does not influence the galaxy-wide assessment of dust emission from the extended quasar host.

\begin{figure*}
\begin{centering}
\includegraphics[width=0.85\textwidth]{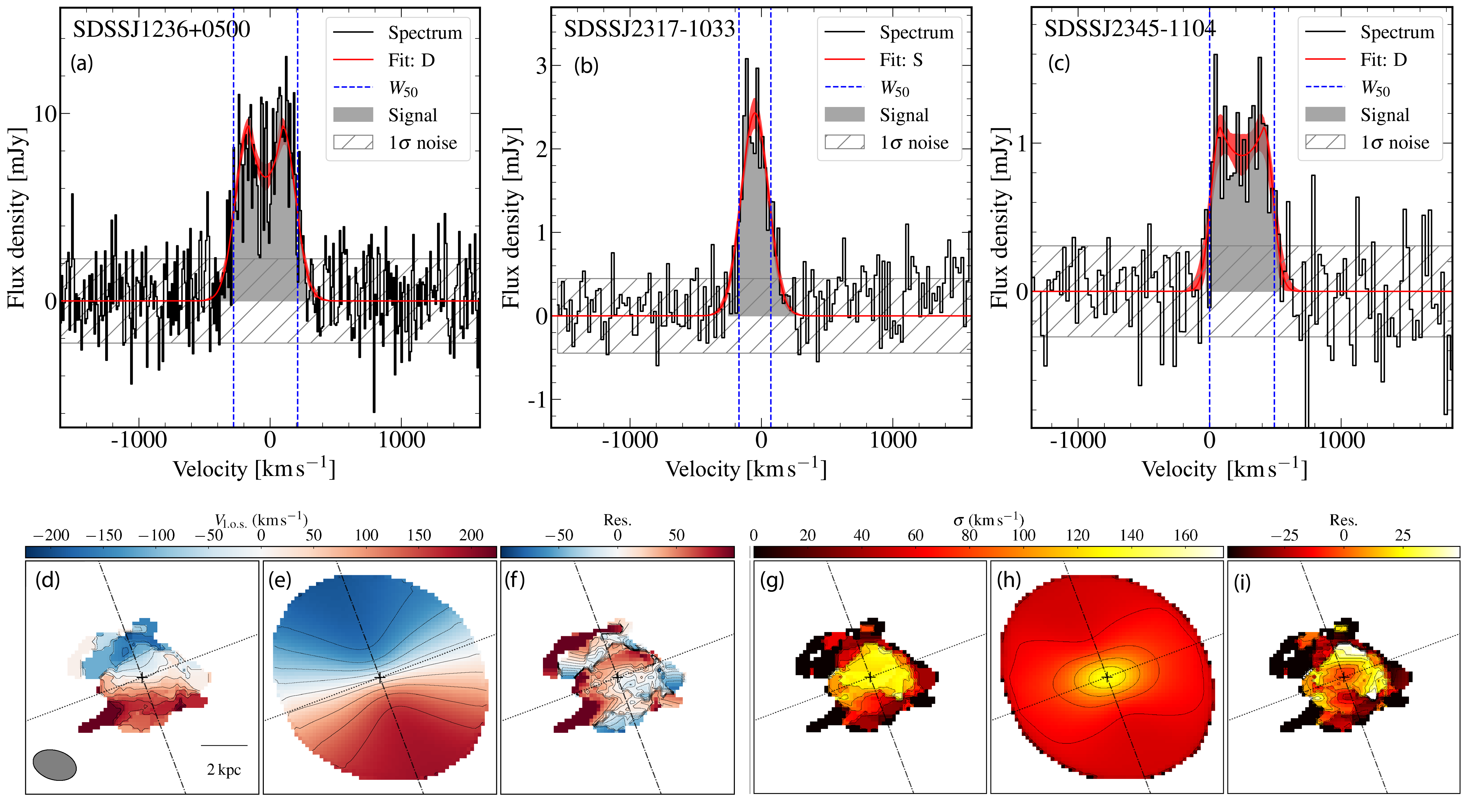}
\end{centering}
\caption{CO (J=5--4) gas kinematics. $top$ Line profiles for SDSS J1236+0500 (a), SDSS J2317-1033 (b), and SDSS J2345-1104 (c) are shown. $Bottom$ Emission-line modeling of SDSS J1136+0500 with best-fit results from 3D-Barolo. The first three panels show the observed velocity map (d), best-fit model (e) and residual emission (f). The three right-most panels (g-i) are the equivalent for the velocity dispersion. All panels are consistent with a rotation-dominated (v/$\sigma$=6.5) disk.}
\label{fig:co_kinematics}
\end{figure*}

\subsubsection{Gas kinematics}
\label{sec:kinematics}

We determine the kinematics of the ISM for SDSS J1236+0500, afforded by the high signal-to-noise detection of CO (J=5-4). Coherent disk rotation may indicate ongoing secular processes driving gas inwards to fuel the quasar. Any disturbances may be an indication for the influence of mergers and/or outflows driven by either the starburst or quasar-mode feedback. However, as shown in Figure~\ref{fig:model_fit_J1236}, it is surprising that the molecular gas in this extreme starburst is so smooth \citep[e.g.,][]{Rizzo2023,Lelli2023} while dynamical instabilities are needed to drive gas to the nuclear region to fuel the luminous quasar.

In Figure~\ref{fig:co_kinematics}, the CO (J=5--4) line has a double-horn profile, characteristics of a rotating disk. The line is symmetric and well fit with a double Gaussian profile shown in red. For a proper assessment, we fit the cube with a thin disk model using 3D-Barolo \citep{DiTeodoro2015}. The individual lower panels show the velocity field, best-fit model, and residual image with the equivalent for the dispersion shown in the three right-most panels. The disk is rotation dominated with $v/\sigma=6.5$ ($v_{rot}=260$ km s$^{-1}$ and $\sigma=40$ km s$^{-1}$). These values are typical for star-forming galaxies at $z\sim2$ \citep{Rizzo2023}. Therefore, there are no signs of ongoing merging or other disturbances on the gas. This is quite remarkable since the galaxy is simultaneously undergoing an intense starburst and luminous quasar phase.

\begin{figure*}
\begin{centering}
\includegraphics[width=0.9\textwidth]{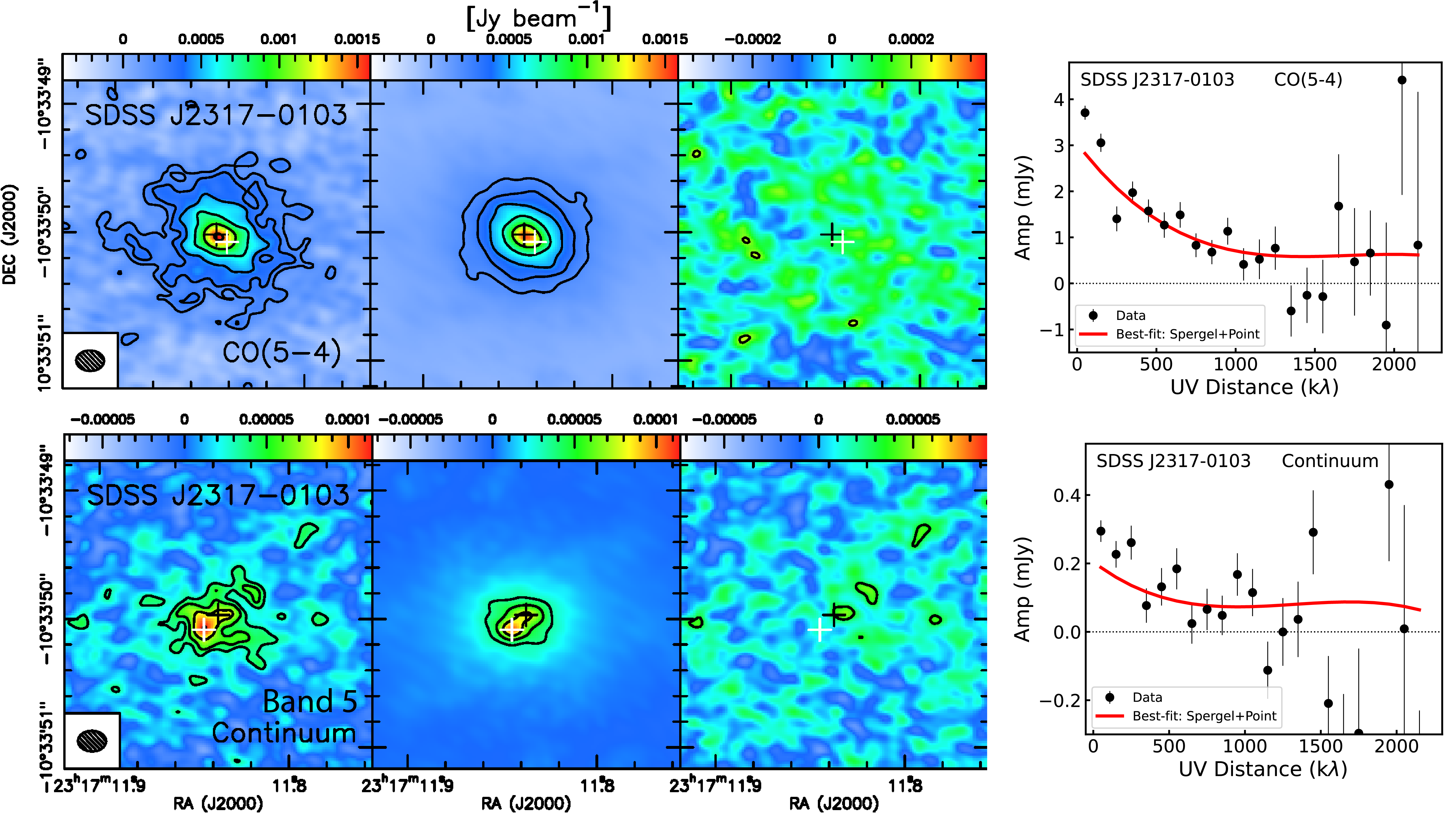}
\end{centering}
\caption{Same as Fig.~\ref{fig:model_fit_J1236} for SDSS J2317-1033 except the 1$\sigma$
rms noise levels for the CO(5-4) and continuum maps are 66 $\mu$Jy beam$^{-1}$ and 11 $\mu$Jy beam$^{-1}$ based on velocity widths reported in Table~\ref{tab:band5_9}.}
\label{fig:model_fit_J2317}
\end{figure*}

\begin{figure*}
\begin{centering}
\includegraphics[width=0.9\textwidth]{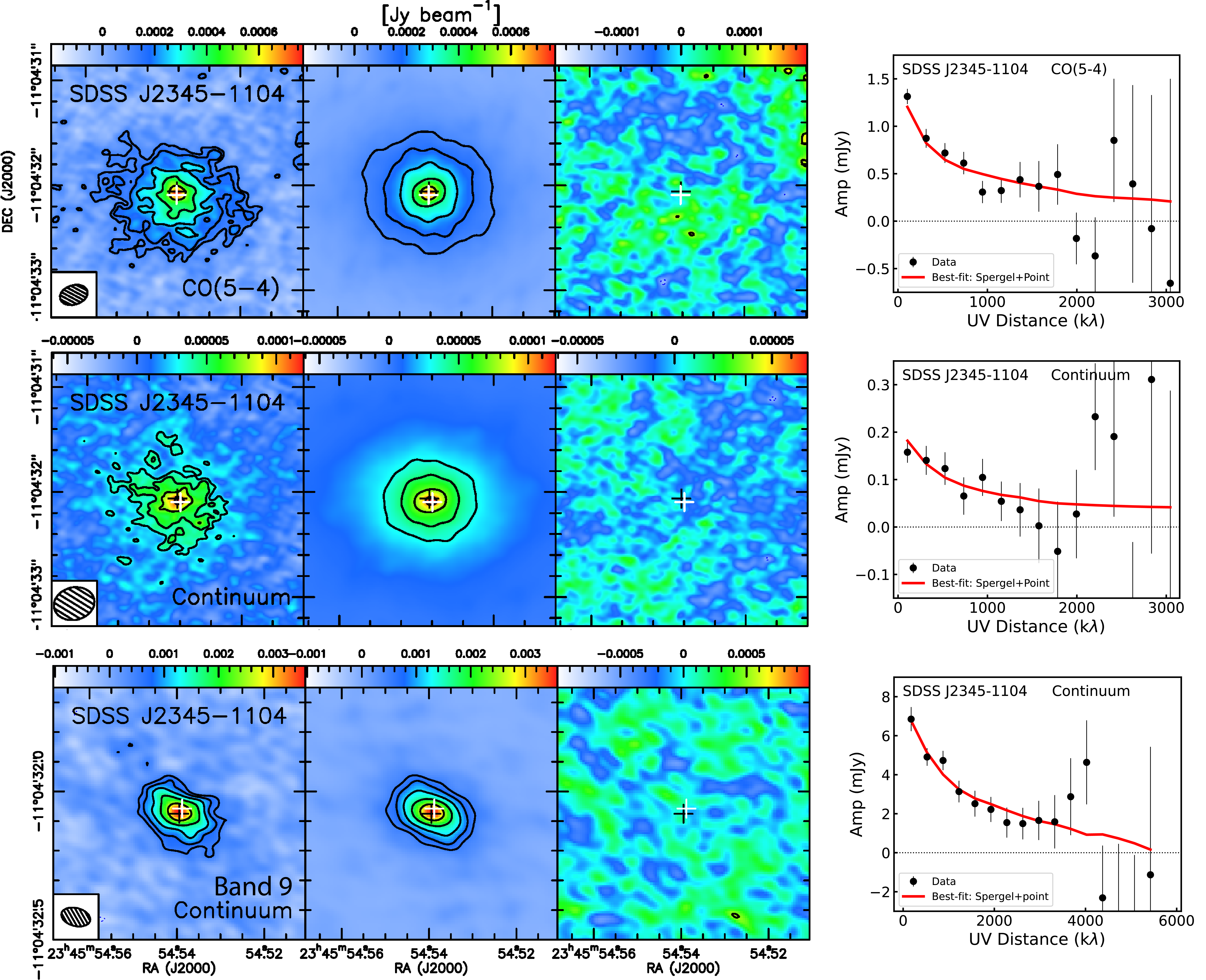}
\end{centering}
\caption{Same as Fig.~\ref{fig:model_fit_J1236} for SDSS J2345-1104 except he 1$\sigma$
rms noise levels for the CO(5-4) and continuum maps are 37 $\mu$Jy beam$^{-1}$, 11 $\mu$Jy beam$^{-1}$ (Band 5) and 0.17 mJy beam$^{-1}$ (Band 9) based on velocity widths reported in Table~\ref{tab:band5_9}.}
\label{fig:model_fit_J2345}
\end{figure*}

\subsection{Luminous quasars in typical star-forming hosts}

SDSS J2317-1033 and SDSS J2345-1104 were targeted as having Band 7 continuum fluxes consistent with lower levels of star formation as compared to the previous case. The band 7 continuum fluxes resulted in estimates of the SFR at levels similar to massive star-forming galaxies on or close to the main-sequence at $z\sim2$ \citep{Schulze2019}.

In Figures~\ref{fig:model_fit_J2317} and ~\ref{fig:model_fit_J2345}, model fits in the $uv$-plane are shown for both CO (J=5--4) and continuum data of each quasar. Both exhibit clearly extended and nearly symmetric CO emission with a peak centered on the location of the quasar. The spatial distribution is well fit by a two-component model as done for SDSS J1236+0500 as evident by the lack of any significant residuals. In both cases, the extended galaxy component contributes to the majority of the CO flux (87.1\% in J2317-1033 and at least 69\% in SDSS J2345-1104). The half-light radii are each similar to 1 kpc as with SDSS J1236+0500. Based on \citet{Tan2024a}, the signal-to-noise of the host is insufficient to accurately determine the Spergel or effective Sersic indices for these two sources. However, we note that the values for both are similar as given in Table~\ref{tab:band5_9}. We conclude that these two quasar hosts have sufficient supplies of molecular gas to form stars at the level seen in the Band 7 data. In Section~\ref{text:sfrs}, we use the ALMA data to explicitly provide new estimates of their star formation rates with both continuum and an inference based on the CO (J=5--4) luminosity.

\subsubsection{Continuum emission}

 We first describe the results (Table~\ref{tab:band5_9}), as done for J1236+0500, of a two-component model fit to the continuum emission for each quasar in band 5 (Figs.~\ref{fig:model_fit_J2317} and~\ref{fig:model_fit_J2345}). Due to the lower S/N detections of the continuum relative to CO, the point-source contributions are not well constrained. However, we run model fits for three different cases: (a) the Spergel index fixed to the output value from the CO fit, (b) index set to -0.3, the mean value for galaxies in the sample of \citet{Tan2024b}, and (3) a free index. For SDSS J2317-1033, the fit for case `a' did not converge while the other two cases result in point source contributions of 44\% and 39\%. These indicate that an unresolved contribution to the continuum emission is present and at a level higher than seen in CO (J=5--4). For J2345-1104, case `a' and `b' provide an upper limit close to 50\%. Case `b' is expressed as an upper limit based on the low S/N of the point-source detection. For a free Spergel index, the constraint is slightly tighter at 33\%. Given these fits, we conclude that there is a high probability for there to be a non-negligible, probably subdominant, unresolved component to the FIR continuum emission. The likely predominance of an extended host galaxy is evident in Figures~\ref{fig:model_fit_J2317} and~\ref{fig:model_fit_J2345} where a rise of the amplitude is seen at short baselines.

With Band 9 observations of J2345-1104 at a beam size of $0.06^{\prime\prime}\times0.10^{\prime\prime}$ and $S/N (beam)=49$ (Fig.~\ref{fig:model_fit_J2345}), the rest-frame continuum emission at 154 $\mu$m is fit with a two-component model where an unresolved source contributes $11.6\%$ ($1.01\pm0.41$ mJy) of the total emission ($8.67\pm1.58$ mJy; Table~\ref{tab:band5_9}). In the lower panels of Figure~\ref{fig:model_fit_J2345}, we show the 2D model fit, residuals and a plot of amplitude versus UV distance. The model has a best-fit effective radius of $0.81\pm0.39$ kpc, which is close to the Band 5 continuum size and within the uncertainties. In this case, the fit converged with a free Spergel index of $-0.71\pm0.06$ which corresponds to a Sersic index likely to be greater than 5, which is indicative of a highly concentrated light profile. This limit is given as a lower bound since there is uncertainty on the precise model parameters (e.g., $r_{eff}$ is constrained at the 2.1$\sigma$ level). Being highly conservative, the unresolved component would contribute at most 26\% of the total emission considering the $3\sigma$ upper bound on its flux ((1.01+3$\times$0.41)/8.67). 

Considering the uncertainties described above, the unresolved emission in both band 5 and 9 is remarkably consistent with expectations from emission from the torus based on the quasar model of \citet{Lyu2017a} as shown in the bottom right panel of Figure~\ref{fig:seds}. Here, we plot $2\sigma$ errors which overlap with the FIR component to the quasar template matched well to the optical/infrared photometry. In the left panel, the total FIR continuum in three bands (5, 7, and 9) is in excellent agreement with a modified black body spectrum with fluxes elevated from the quasar template. Therefore, we reach the same conclusion as J1236+0500 that the FIR emission from J2345-1104 is dominated by the extended host galaxy, thus providing a clean assessment of the ISM of a luminous quasar, even in the case of a SF MS host.

\subsection{CO -- FIR relation}

Considering the results for the three quasars together, we examine the relation between extended FIR and CO (J=5--4) luminosities in the context of published results from a larger study of star-forming galaxies with varying levels of AGN activity \citep{Valentino2021}. Based mainly on low-resolution ALMA observations of X-ray detected AGN ($L_{2-10~keV}=10^{43-45}\,\mathrm{erg}\,\mathrm{s}^{-1}$), these authors find a general lack of strong AGN contribution to the CO (J=5--4) luminosity and a minor contribution to the FIR continuum on the Rayleigh-Jeans tail. For the luminous quasar population, almost completely missed in their work, we should see deviations in the FIR continuum luminosity relative to the CO (J=5--4) emission, if there is additional heating of dust on extended scales due to the quasar.

\begin{figure*}
\includegraphics[width=0.45\textwidth]{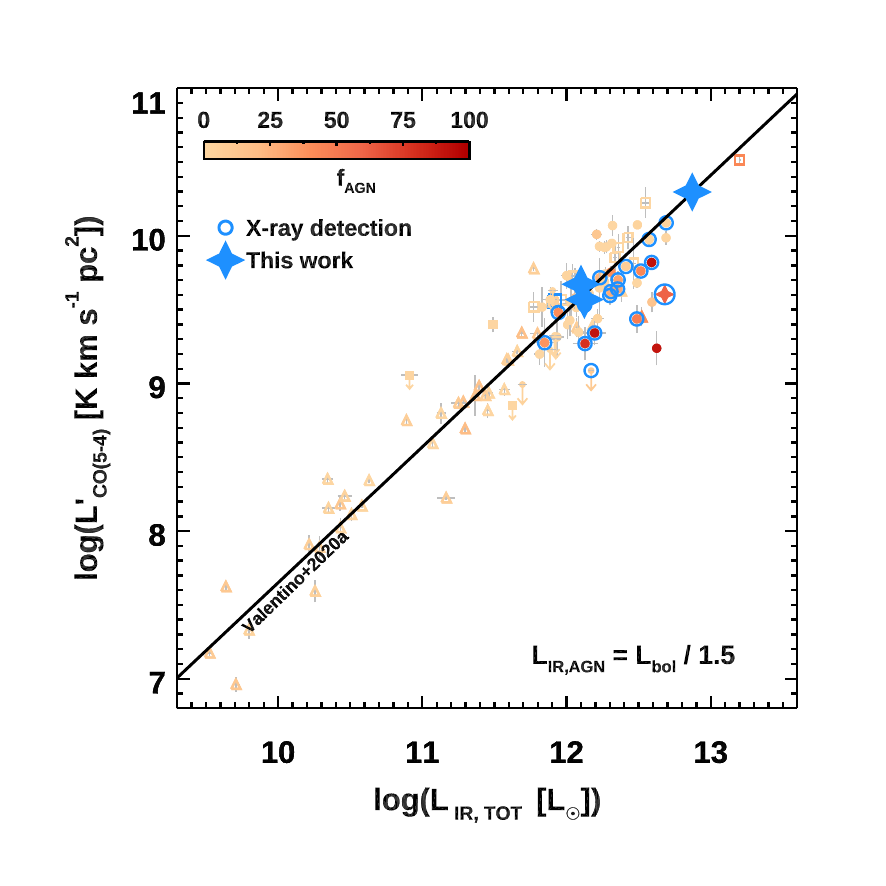}
\includegraphics[width=0.45\textwidth]{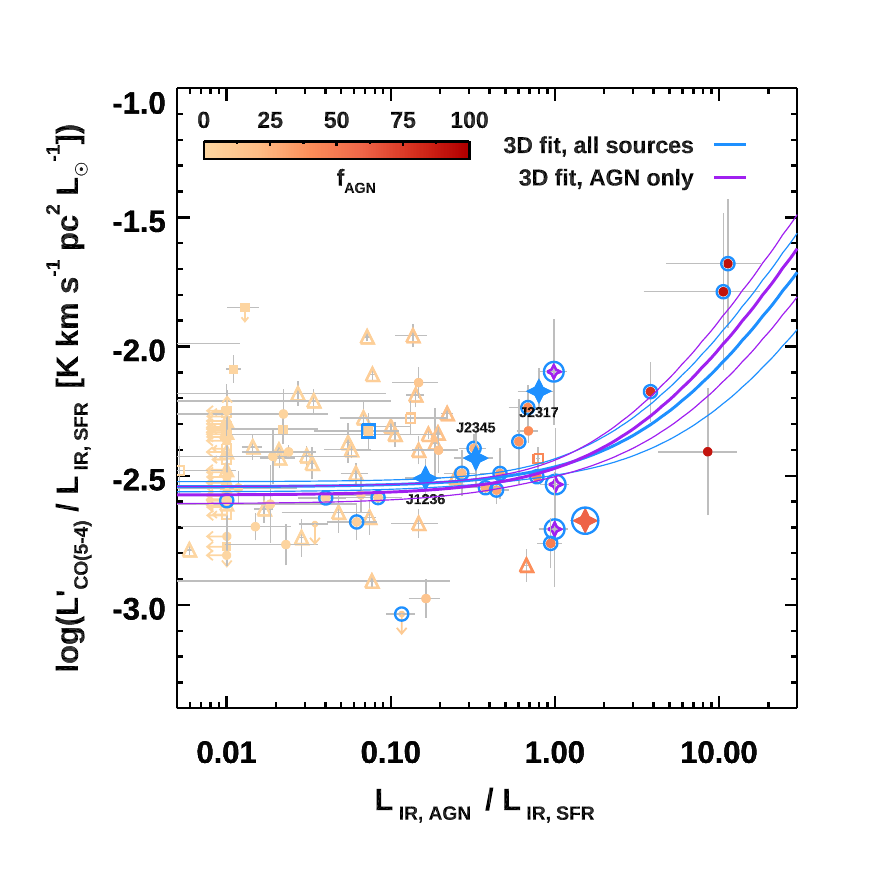}
\caption{$left$: Relation between CO (J=5--4) and total IR luminosities (comprehensive of both the unresolved and resolved components, assigned to the AGN and SF-heated components, respectively) for our three quasars (blue stars) with comparisons to samples at $z\sim1.5-2$ compiled from the literature \citep{Valentino2021} and color coded according to their L$_{IR, AGN}$/L$_{IR, SFR}$ ratio. Hard X-ray detected sources ($L_{2-10~keV}>10^{43}\,\mathrm{erg}\,\mathrm{s}^{-1}$) are shown as open blue circles. Open purple stars are from \citet{Bischetti2021}. The orange filled star marks the object from \citet{Brusa2018}. Open triangles and squares represent local objects and distant SMGs from \citet{Liu2021}. Filled squares indicate the ASPECS sample, with X-ray AGNs marked with blue empty squares \citep{Boogaard2020}. The black solid line shows the relation in \citet{Valentino2020}. $Right$: The same relation, but now showing the ratio of the AGN-to-SF contribution to the IR luminosity (L$_{IR, AGN}$/L$_{IR, SFR}$) and the CO (J=5--4) to L$_{IR, SFR}$ luminosities (L$^{\prime}_{CO (J=5-4)}$/L$_{IR, SFR}$) for the same samples. The blue and purple lines indicate the best-fit multiple linear regression models (and their uncertainties) to the L$^{\prime}_{CO (J=5-4)}$ luminosity as a function of L$_{IR, SFR}$ and L$_{IR, AGN}$ for the overall sample (including our quasars) and to the AGN sample only, respectively. The symbols are the same as in the left panel.}
\label{fig:lsfr_lco}
\end{figure*}

In Figure~\ref{fig:lsfr_lco}, we show the location of our three quasars in the space of L$^{\prime}_{CO (J=5-4)}$, L$_{IR, SFR}$, L$_{IR, AGN}$, and their combined total infrared luminosity (L$_{IR, Tot}$ = L$_{IR, SFR}$+L$_{IR, AGN}$). In both panels, we assumed that the point source emission originates from the unresolved component of the quasar, while the extended component is assigned to the heating solely due to star formation.

We find that our three targets, despite their high bolometric luminosities, are overall consistent with the locus of the star-forming galaxies and low-luminosity AGN in the compilation by \citet{Valentino2021}. Two of our quasar hosts with more typical IR luminosities fall within the locus of the larger comparison sample while the our single starburst adds to the high end which has only one other galaxy at these extreme values. On the one hand, if there were quasar heating of extended dust without powering extra gas heating, one would expect our quasars to lie to the right of the relation, established mainly with inactive galaxies or low-luminosity AGNs (Fig.~\ref{fig:lsfr_lco}; $left$ panel). On the other hand, if the quasar were to contribute to the heating of the gas traced by CO(J=5–4) at fixed dust luminosity, we would expect our quasars to lie above the relation. Neither of these situations are seen.

Figure~\ref{fig:lsfr_lco} ($right$ panel) shows an alternative view of the same parameter space, now highlighting how the ratio of the AGN-to-star-forming contribution to the IR luminosities compares with that of the CO (J=5--4) to L$_{IR, SFR}$ luminosities. Any deviation from the correlation between the two quantities due to the quasar contribution should appear as vertical offsets on the right-hand side of the plot. The three quasars analyzed in this work are consistent with the locus spanned by the literature sample, even if lying slightly above the best-fit relation (Section \ref{text:xdr}).  We conclude that the quasar contribution to the CO (J=5--4) emission is lesser, and the FIR luminosity is dominated by star formation, supporting the hypothesis that the prediction of the infrared luminosity from CO (J=5-4) is in agreement with the measured FIR continuum on galaxy-wide scales. In Section~\ref{text:xdr}, we evaluate and quantify the possible quasar contribution to the CO (J=5-4) emission.

\subsubsection{Quasar influence on CO (J=5-4)}
\label{text:xdr}

Studies show that the AGN can contribute to CO excitation of \hbox{high-$J$} transitions, particularly from an X-ray Dominated Region\citep[XDR;][]{Vallini2019,Pensabene2021,Wolfire2022}. However, for $J=5$, the effect is not as strong as for higher-$J$ transitions, and the emission, even at the highest QSO luminosities, might still be dominated by processes related to star formation. Following the simplified empirical approach in \citet{Valentino2021}, we attempt to quantify the effect of the AGN on the CO (J=5-4) emission in our quasars by modeling the CO gas luminosities as a function simultaneously of the extended FIR emission (attributed to the heating due to star formation in the disk) and the unresolved AGN component (entirely assigned to the central nuclear activity). We understand that there may be compact components due to star formation which may not have been accounted for in the model of the extended component. Thus, our results are likely to be an upper limit on the quasar contribution.

The best-fit models and their uncertainties, calculated including our targets in the literature compilation, are shown in Figure~\ref{fig:lsfr_lco} ($right$) where an upturn in
$L'_{\rm CO(5-4)}$ appears when $L_{IR,AGN}>L_{IR, SFR}$. The best-fit coefficients are similar to those retrieved in
\citet{Valentino2021}. Individually, we find coefficients of
$0.12\pm0.04$, $0.12\pm0.03$ and $0.30\pm0.24$ for J1236+0500, J2317-1033 and J2345-1104, respectively. Given the large uncertainty on the estimate for J2354-1104, we choose to use the value of $0.12\pm0.04$ as our best assessment on the quasar contribution to CO (J=5-4) for a given IR luminosity among our bright QSO targets. We conclude that there is no evidence for these quasars influencing the CO (5-4) luminosity in such a way as to impact its use as a proxy for the SFR. Our results are in agreement with low-redshift studies of lower luminosity AGNs \citep{Esposito2022,Esposito2024}.

\subsection{Presence of star formation in luminous quasar hosts}
\label{text:sfrs}

\begin{figure}
\includegraphics[width=0.45\textwidth]{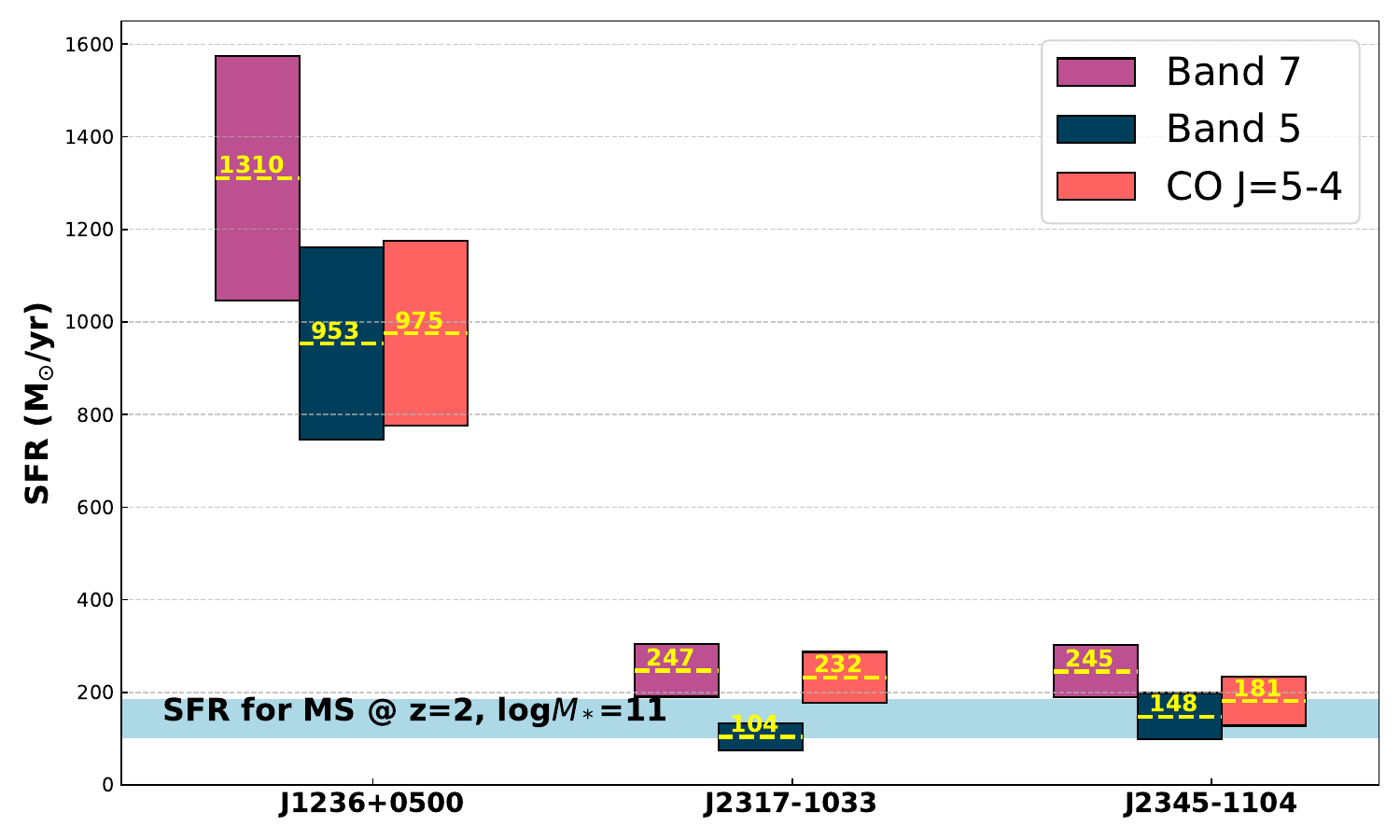}
\caption{SFRs for our three luminous SDSS quasars. The methods are shown for (1) template fitting to the FIR continuum emission using the unresolved Band 7 data or the resolved Band 5 data (minus the quasar contributon for J2345-1104), and  (2) that inferred from the total CO J=5--4 emission-line flux for each target. Mean SFRs for each method are shown in yellow. The grey band highlights the typical SFR for massive (log M$_{*}$=11) MS galaxies at $z=2$.}
\label{fig:sfr_comp}
\end{figure}

Inferring the SFRs using the CO (J=5--4) luminosities of the extended components for the three quasars and the CO (J=5--4) -- L$_{IR}$ relation \citep{Daddi2015} with its uncertainty, the total infrared luminosities of 7.3, 1.5 and 1.35 (units of 10$^{12}$ L$_{\odot}$) are found and correspond to SFRs of 975$\pm200$,  232$\pm55$ and 181$\pm53$ M$_{\odot}$ yr$^{-1}$ using the relation in \citet{Kennicutt2012}. In Figure~\ref{fig:sfr_comp}, the CO-based SFRs do have differences with those based on the previously reported Band 7 continuum luminosities \citep{Schulze2019}. While there is a reduction in the SFR for J1236+0500 by 335 M$_{\odot}$ yr$^{-1}$, the CO-based estimate does not alter its classification as a starburst galaxy, well above the MS. The CO-based SFRs for the other two cases are very close to the MS, if their host galaxies have $\sim10^{11}$ M$_{*}$ at $z=2$. In Section~\ref{sec:dynmass}, we assess whether this is the appropriate mass range for the hosts to these three quasars. Observations with JWST may be able to discern the stellar mass of their host galaxies \citep{LiHoChen2025} while their optical emission is probably too bright for HST.

Then we compare the SFRs based on the FIR continuum luminosity with the CO (J=5--4) luminosity in Figure~\ref{fig:sfr_comp}. The SFRs from the FIR continuum are shown for the resolved Band 5 and unresolved Band 7 separately. An AGN contribution has been removed for those with a $>$3$\sigma$ detection of a central unresolved component for the Band 5 data; this pertains to the SFR based on the continuum for SDSS J2345-1104. For the upper limits on the AGN contribution, we include the reduction in SFR that considers a subtraction of the maximal contribution from the AGN in Table ~\ref{tab:band5_9}. We do not use the Band 9 fluxes here since we do not have such data for all three quasars. As shown, the SFRs based on the high resolution continuum emission in band 5 are significantly reduced by factors of 27, 58 and 61\% from their band 7 estimates for J1236+0500, J2317-1033 and J2345-1104 respectively. In all three cases, the CO (J=5--4) estimates are higher than the band-5 continuum estimates by $1-2\sigma$. The band 5 continuum estimates for J1236+0500 and J2345-1104 are in excellent agreement with the CO-based SFRs. Even with differences in SFRs for these indicators, the original classification of J1236+0500 as an extreme starburst and the other two (J2317-1033, J2345-1104) having SFRs consistent with the MS holds as reported in \citet{Schulze2019}. We caution that unresolved continuum-based SFRs may be impacted by unresolved torus emission as described above; however, the level of emission is subdominant to the dust heating from ongoing star formation on the extended scales of the host. For J1236+0500 and J2345-1104, the band 9 observations lend support to this conclusion by demonstrating the existence of a torus component, in agreement with the decomposed Band 5 fluxes as shown in Figure~\ref{fig:seds}, at a level below that of the extended thermal emission.

\section{Discussion}

\subsection{AGN contribution to the FIR continuum emission guided by simulations}

Based on radiative transfer models, \citet{McKinney2021} predict an enhancement in the thermal emission from a galaxy hosting an AGN based on a simulation of a merger at $z\sim2$, compared to one without an AGN. Their model includes both the hot torus and the more extended dust on the scale of the host galaxy. As the bolometric luminosity of the model AGN reaches up to 10$\times$ the bolometric stellar luminosity, the FIR luminosity at 160 (500) $\mu$m is boosted by 5.2 (2.4)$\times$, relative to the case with an inactive AGN as reported in Table 1 of their study. These factors represent the case of a maximal boost which is likely appropriate for our highly luminous quasars. 

As presented above, our observations do not support the above theoretical scenario. For all three quasars, the strength and distribution of the extended FIR continuum is very similar to that predicted by the CO (J=5-4) data. This is clearly indicative of high levels of molecular gas and dust that is forming stars thus supporting the interpretation of the extended dust emission being predominantely heated by stars rather than a luminous quasar.

\begin{figure}
\includegraphics[width=0.45\textwidth]{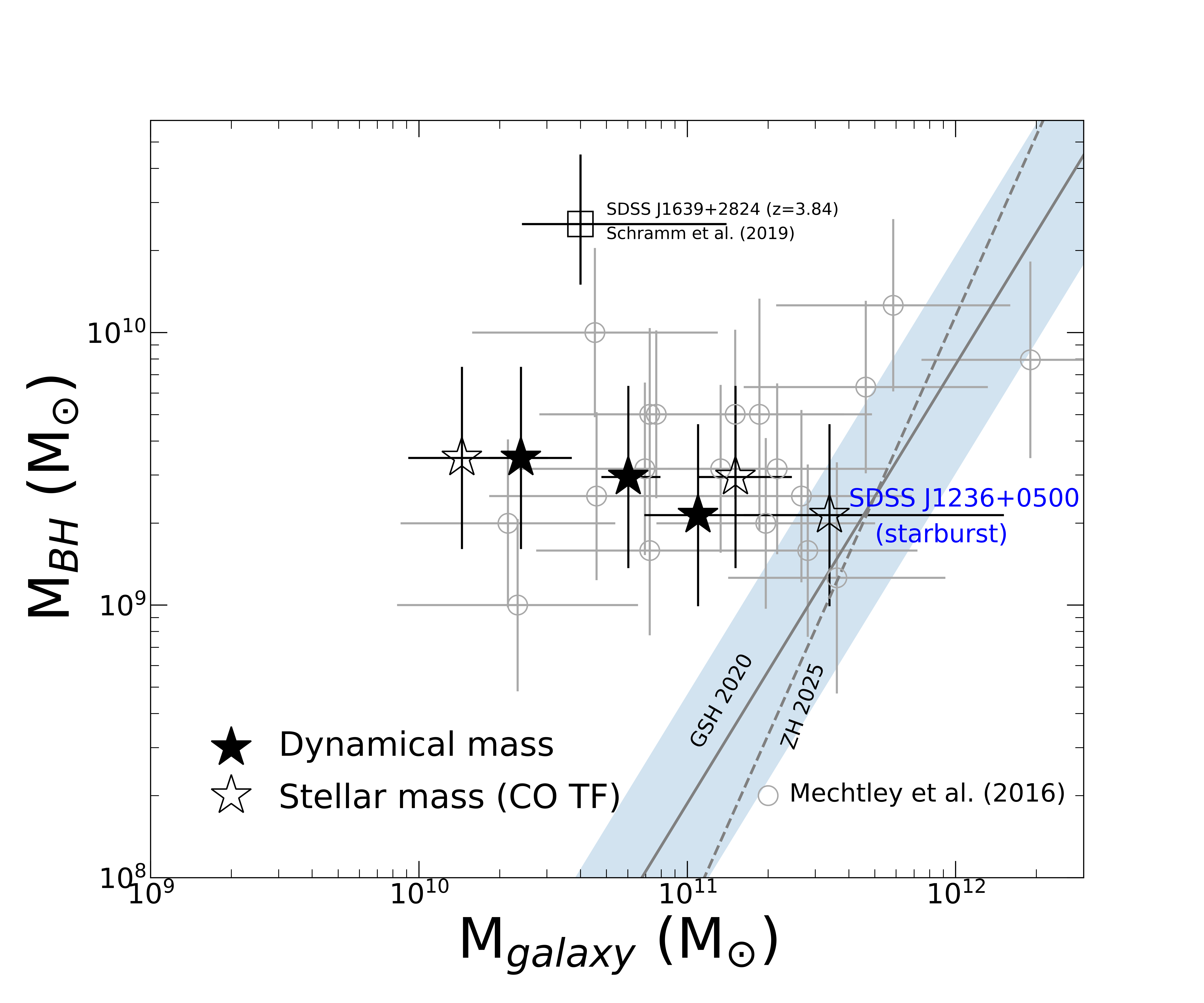}
\includegraphics[width=0.45\textwidth]{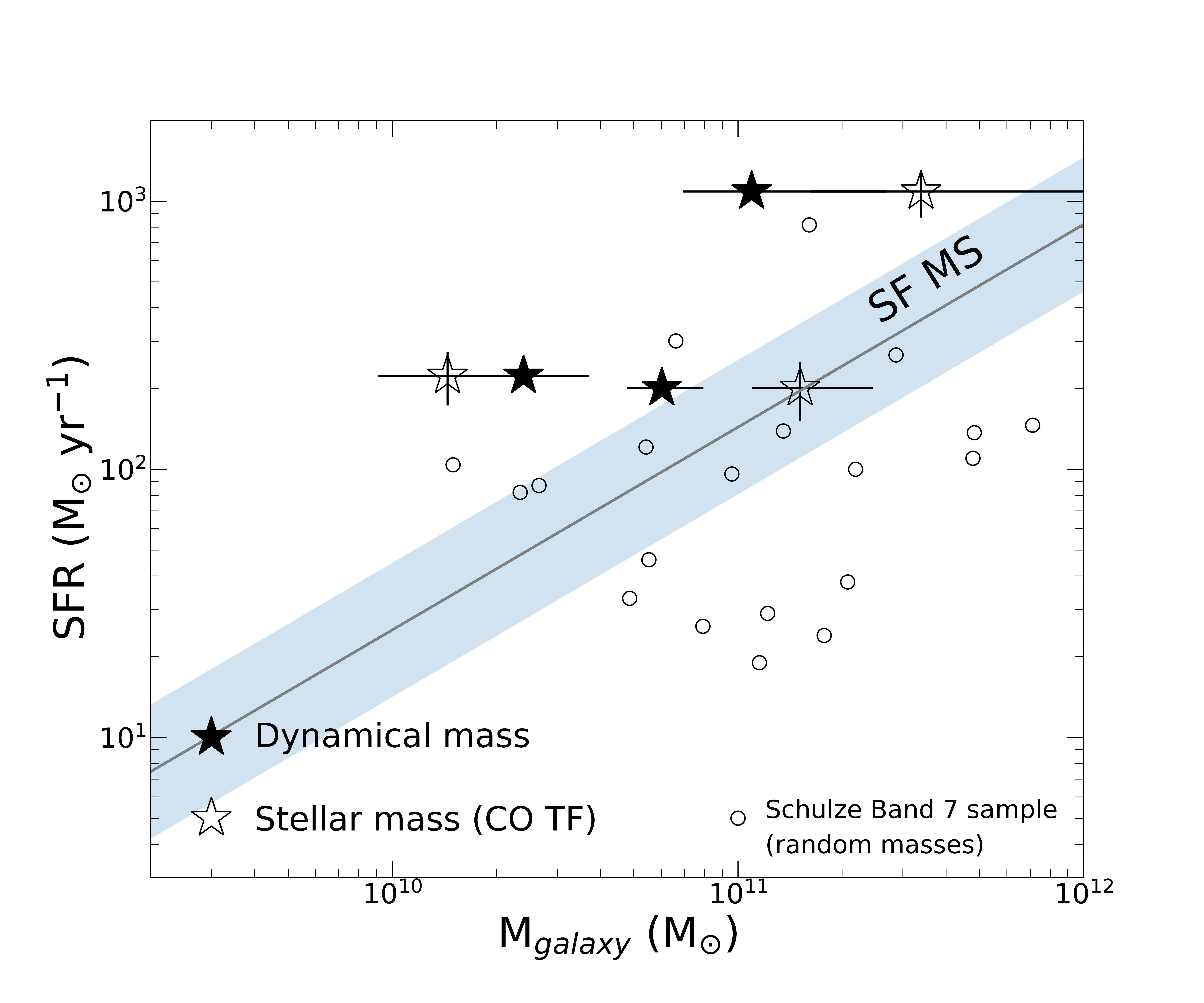}
\caption{(a) Black hole mass versus galaxy mass where the latter is either the dynamical (filled star) or stellar (open star; CO Tully- Fisher; \citealt{Tiley2016a}) mass with each based on the CO J=5--4 emission line. For reference, we plot the values from the $z\sim2$ quasar sample of \citet{Mechtley2016}, the most extreme offset quasar (square; \citealt{Schramm2019}), and the local mass relations of inactive \citep{Greene2020} and active (type 1 AGNs; \citealt{Zhuang2023}) including its mean and dispersion for the former. (b) Star formation rate (based on CO J=5--4) versus galaxy mass as described for the top panel. For reference, the average SF MS at z=2 \citep{Speagle2014} is shown by the slanted line and shaded region. For illustrative purposes, the full quasar sample from \citet{Schulze2019} with FIR continuum measurements (i.e., SFRs) is plotted (open grey circles) with stellar mass estimates randomly assigned based on the stellar mass distribution of the \citet{Mechtley2016} sample.}
\label{fig:ms}
\end{figure}

\subsection{Dynamical and inferred stellar masses}
\label{sec:dynmass}

We attempt to validate the assertion in Section~\ref{text:sfrs} that the stellar masses of our quasar host galaxies are expected to be $\sim10^{11}$ M$_{\odot}$ for comparison of the SFRs for our three quasars to that of inactive star-forming galaxies. We would expect their host galaxies to be this massive given the high mass of their black holes. To do so, we fit the local CO J=5--4 emission lines from our three quasars to estimate their dynamical mass (and stellar mass using the local CO Tully-Fischer relation), as done in \citet{Ho2007}, since direct detection of the stellar emission is not currently available and is likely to be challenging given the brightness of our quasars, possibly even with JWST. 

From the reduced ALMA data cubes, one-dimensional spectra are extracted using pixels within a 1$^{\prime\prime}$ diameter aperture, centered on each target, and having S/N $>$ 2. For J1236+0500 and J2345-1104, a double-horn profile is an appropriate choice for the model and expressed as a function of frequency as follows: 
    \begin{align}
        f(v) = 
        \begin{cases}
            A_{\rm G} \times \exp{\frac{-[v-(v_0-w)]^2}{2\sigma^2}} \geqslant &v<v_0-w \\
            A_{\rm C}+ a(v-v_0)^2 &v_0-w\leq v\leq v_0+w \\
            A_{\rm G} \times \exp{\frac{-[v-(v_0+w)]^2}{2\sigma^2}} \geqslant &v>v_0+w \\
        \end{cases}
    \end{align}

\noindent Here, the observed line center is represented by $\nu_{0}$ and the half-width is given as $w$ which is expressed as $W_{50}$, the full-width at half the maximum value. For J2317-1033, we fit the line with a single Gaussian profile rather than a double-horn function since the spectrum is narrow with only one peak.

Using these best-fit models, we assume that the gas distribution is described by a rotation-dominated, thin disk \citep{Neeleman2021} thus the dynamical mass enclosed within a radius, $R$, is: 

\begin{equation}
M_{\rm dyn} (R) = \frac{v_{\rm circ}^2}{G}R = 2.355\times 10^5v_{\rm circ}^2 R
\end{equation}

\begin{equation}
v_{\rm circ} = 0.5\times W_{\rm 20}/\sin i
\end{equation}

\begin{equation}
cos~i = \frac{b}{a}
\end{equation}

\noindent where $v_{\rm circ}$ is the circular velocity in units of $\rm km\,s^{-1}$. The inclination ($i$) of the galaxy is estimated from minor-to-major axis ratio. The dynamical mass estimates with uncertainties are based on a Markov Chain Monte Carlo (MCMC) approach.

Lastly, we use the local CO Tully-Fisher relation \citep{Tiley2016a} to infer the stellar mass based on the COLD-GASS survey:

    \begin{align}
        \log M_\star/M_\odot = 3.3\pm 0.3 \left(\log \frac{W_{50}/\sin i}{\rm km\,s^{-1}} - 2.58\right) + 10.51\pm0.04,
    \end{align}

\noindent However, \cite{Tiley2016b} indicates that the zero-point of Tully-Fisher relation can have a $-$0.41 dex offset over the last 8 billion years ($z\sim1$). Since our targets are $z\sim2$ quasars, the stellar mass can be smaller than that based on the local relation for which we use in any case.

In Figure~\ref{fig:ms}, we plot the results of this exercise. In panel a, the stellar mass is shown along with their black holes masses. As evident, the stellar masses from the CO TF relation are close to $10^{11}$ M$_\odot$ in two cases while the dynamical masses are shifted to lower values. For reference, we include the sample of quasars at $z\sim2$ from the \citet{Mechtley2016} study as a reference which generally cover the region of our three quasars. Therefore, we claim that the host galaxy masses are consistent to those expected based on our knowledge of similar efforts at $z=2$. We point out that J1236+0500 with a starbursting host is on the local mass scaling relation, considering the inferred stellar mass from CO. This agrees with possible pathways for (starburst) galaxies to migrate onto the local relations \citep[e.g.,][]{Zhuang2023}.  

We highlight that quasars offset from the local mass scaling relation are expected since selection effects (and measurement uncertainties) are a strong determinant in the observed location of quasar-selected samples in the $M_{BH}-M_*$ plane at high redshift \citep[e.g.,][]{Schulze2011,Li2025}. However, some published cases are evidently so extreme that their offsets from the local relation cannot be explained simply by measurement uncertainties. For example, \citet{Schramm2019} find that one of the most massive black holes in the universe at $z=3.84$ is offset by over an order-of-magnitude (square symbol in Fig.~\ref{fig:ms}) based on Subaru K-band imaging (0.2$^{\prime\prime}$) with adaptive optics and high resolution ALMA CO (J=4-3; beam size of $0.19^{\prime\prime}\times 0.13^{\prime\prime}$) observations, which provides a dynamical mass estimate. Such offset AGNs are being found at much lower black hole ($M_{BH}\sim10^8$ M$_\odot$) and galaxy mass ($M_{*}\sim10^9$ M$_\odot$) with JWST \citep[e.g.,][]{Juodzbalis2024,Mezcua2024}, which lend support for a heavy seed channel in black hole formation \citep{Volonteri2005,Natarajan2024} for these individual cases.

Regarding our main interest, we plot the SFR versus galaxy mass in panel b. While the galaxy masses significantly shift closer to or away from the SF MS at $z=2$ depending on the method, the SFRs never fall below the MS. The hosts for our three quasars are forming stars either at rates consistent with the MS or higher, if the stellar masses are lower than $10^{11}$ M$_\odot$.

\subsection{Implications for quasar-mode feedback}

Hydrodynamic simulations require energy injection from SMBHs to regulate star formation thus matching the mass function of galaxies and relation to dark matter halos. Many studies to date have searched for signs of the quenching of star formation in quasar and AGN hosts galaxies with mixed results. In fact, overwhelmingly, the host galaxies of quasars and luminous AGNs are preferentially star-forming over a wide range of redshift. This is further supported by this study with star formation rates at the level of MS galaxies at $z=2$ or higher.

Rather than looking for such an obvious smoking gun of quasar-mode feedback, the additional FIR continuum emission, seen as an unresolved central source, may represent heating from the quasar. We are motivated by the recent results from \citet{Tsukui2023} who have found a second warmer thermal component in a luminous infrared galaxy at z=4.4 which is attributed to AGN heating at a level which reduces the SFR estimate by a factor of two, without altering its nature as a starburst. Interestingly, as reported above, the unresolved components have FIR luminosities of 0.2, 1.6 and 4\% of the bolometric luminosity for each quasar. If we attribute this solely to quasar-heated dust, these represent an amount of energy roughly similar to the 5\% of energy deposited to the ISM in hydrodynamic simulations \citep{DiMatteo2005,Weinberger2017}. The fact that we do not see galaxy-scale outflows (Section~\ref{sec:kinematics}) in J1236+0500 may indicate that the heating is indeed a nuclear phenomenon which may explain the lack of a consensus regarding molecular gas content and SFRs on galaxy wide scales. However, this quasar heated dust is likely in the form of the well-known torus rather than more widely spread on galactic scales thus limiting such support for quasar heating models.

\section{Summary}

We have carried out high resolution (beam size of $\sim0.2^{\prime\prime}$) ALMA observations of the FIR continuum and CO (J=5--4) for three of the most luminous quasars at $z=2$ to determine whether there is any impact of the quasar on the ISM on the extended scale of the host galaxy and in the central kiloparsec. Band 9 observations for two quasars at 0.3$^{\prime\prime}$ and 0.06$^{\prime\prime}$ spatial resolution are instrumental in our overall assessment. The motivation to determine whether the FIR continuum is an effective indicator of ongoing star formation comes from results of merger simulations \citep{McKinney2021}, which indicate that central and extended (galaxy wide) dust emission can be heated by a quasar. If realized, this effect can hamper the use of the FIR continuum as a SFR tracer. These results present a simple observational test which can be carried out with high-resolution ALMA data with an answer that sheds light on quasar-mode feedback and regulation of the growth of galaxies. 

With the high-resolution ALMA observations, we fit the data in $uv$-space using two model components to represent an unresolved point source and an extended galaxy based on a Spergel light profile. The continuum and line emission are primarily dominated by an extended component, characteristic of the ISM in massive star-forming galaxies. Fainter unresolved components are detected; however, their contribution to the total continuum and line emission is inconsequential in assessing the global ISM properties of quasar hosts. To support this claim, we specifically find the following by first examining the high S/N data on a starbursting quasar host (J1236+0500) and then expanding the analysis to two SF MS hosts (J2317-1033, J2345-1104):\\

\begin{itemize}

    \item The FIR continuum is well modeled with an extended, smooth, and compact ($r_{eff}\sim1$ kpc) profile resembling a disk in both Band 5 and 9. 
        
    \item Central, unresolved continuum components are detected with a minority contribution to the total continuum emission: 12\% or less in band 9 (for two quasars) and $<44\%$ in Band 5 for the third quasar. The unresolved continuum flux is consistent with expectations from a torus based on the standard quasar SED template of \citet{Lyu2017b}.

    \item Luminous CO (J=5--4) emission is present and follows a similar distribution to the continuum.  The high level of CO emission does not support a scenario where the quasar is responsible for heating the extended dust above that from star formation. If so, there should be a lack of co-spatial CO emission, for which is not observed.
    
    \item The kinematics of J1236+0500 is modeled using 3D-Barolo, which shows properties of a normal, undisturbed disk in rotation (v/$\sigma$=6.5).
    
    \item A central, unresolved ($<$1 kpc) component to the CO (J=5--4) emission is detected in two cases (5.7\% and 12.9\%) and subdominant in the third case, which are the likely result of a minor AGN contribution to CO excitation.

    \item With unresolved components in both CO (J=5-4) and continuum we are able to quantify the relation between the two, which nearly agrees with analysis based on a larger sample having lower resolution data \citep{Valentino2021}.

\end{itemize}

Taking the above into consideration, the FIR continuum and CO (J=5-4) are both effective at providing an assessment of ongoing star formation, essentially uncontaminated, in the host of luminous quasars and indicate that such star formation is present. Therefore, these results support measurements of SFRs from single-band FIR continuum estimates with lower resolution data \citep{Stanley2017,Schulze2019}, the average location of quasar hosts on the SF MS and a lack of any sign of quasar-mode feedback. These results further highlight the need for more effort at physical scales below a kpc with ALMA (and JWST) to reveal the influence of black hole accretion on central gas/dust reservoirs and star formation.

\section*{Acknowledgements}

This work has made use of data from the European Space Agency (ESA) mission
{\it Gaia} (\url{https://www.cosmos.esa.int/gaia}), processed by the {\it Gaia}
Data Processing and Analysis Consortium (DPAC,
\url{https://www.cosmos.esa.int/web/gaia/dpac/consortium}). Funding for the DPAC
has been provided by national institutions, in particular the institutions
participating in the {\it Gaia} Multilateral Agreement. LCH was supported by the National Science Foundation of China (12233001), the National Key R\&D Program of China (2022YFF0503401), and the China Manned Space Program (CMS-CSST-2025-A09).

\bibliographystyle{mnras}
\bibliography{jdsrefs}{} 



\begin{table*}
\centering
\caption{Quasar targets at $z=2$ and previous ALMA Band 7 results}
\label{tab:sample}
\begin{tabular}{llcccccc}
\hline
SDSS ID&z&log~M$_{BH}$&log~L$_{bol}$&S$_{850\mu m}$&log~L$_{850\mu m}$&SFR$_{temp}$&AGN\\
&&($M_{\odot}$)&(erg s$^{-1}$)&(mJy)&(erg s$^{-1}$)&($M_{\odot}$ yr$^{-1}$)&contribution\\
\hline
SDSSJ123649.43+050023.3&1.941&9.33&47.0&7.04$\pm$0.09&44.82&815$_{-231}^{+221}$&$\sim$2\%\\
SDSSJ231711.83--103349.9&2.004&9.54&47.1&1.47$\pm$0.10&44.17&146$_{-41}^{+39}$&$\sim$17\%\\
SDSSJ234554.54--110432.0&1.948&9.47&46.9&1.34$\pm$0.11&44.04&139$_{-39}^{+38}$&$\sim$13\%\\
\hline
\end{tabular}
\end{table*}

\begin{table*}
\centering
\caption{ALMA measurements (Band 5, 7, and 9)}
\begin{threeparttable}
\label{tab:band5_9}
\begin{tabular}{lcccc}
\hline
Quantity&J1236+0500&J2317--1033&J2345--1104&Units\\
\hline
$z_{CO}$&1.9470&2.0037&1.9496&Redshift\\
S$_{CO J=5-4}$ (host)&5.82$\pm$0.26&2.71$\pm$0.34&1.02$\pm$0.22& mJy\\
S$_{CO J=5-4}$(point source)&0.35$\pm$0.08&0.40$\pm$0.0.09&0.28$\pm$0.15& mJy\\
\% point source$^c$&5.7&12.9&$<$31 (3$\sigma$)&\%\\
$\nu_{Spergel}^a$&$0.34\pm0.43$&$-0.71\pm0.06$&$-0.62\pm0.19$\\
$\Delta$v (FWHM)&455&220&480&km s$^{-1}$\\
L$^{\prime}_{CO J=5-4}$&1.98e10&0.47e10&0.37e10&K km s$^{-1}$ pc$^{2}$\\
$r_{eff, host}$&1.01$\pm$0.08&1.17$\pm$0.33&1.09$\pm$0.50&kpc\\
$L_{TIR}$ (CO J=5--4)$^e$&7.3&1.5&1.35&$10^{12}$~$L_{\odot}$\\
SFR (CO J=5--4)&975$\pm$200&232$\pm$55&181$\pm$53&$M_{\odot}$ yr$^{-1}$\\
\hline
S$_{Band~5}$ (host)$^b$&0.92$\pm$0.08 (0.98, 0.70) & --- (0.10$\pm$0.02, 0.11)&0.15$\pm$0.04 (0.15, 0.14)&mJy\\
S$_{Band~5}$ (point source)$^b$&0.15$\pm$0.03 (0.11, 0.14)& --- (0.08$\pm$0.01, 0.07)&0.05$\pm$0.03 (0.04, 0.07)&mJy\\
$\%$ point source$^c$&14.2 (10.1, 16.8)&--- (44, 39)& $<$ 50 ($<$ 47, 35)& \%\\
$r_{eff, host}$&1.64$\pm$0.23&0.72$\pm$0.32&$0.92\pm$0.48&kpc\\
$L_{TIR}$ (Band 5)$^e$&6.4&0.7&1.0&$10^{12}$~$L_{\odot}$ (8-1000 $\mu$m)\\
SFR (Band 5)&953$\pm$208&104$\pm$29&149$\pm$50&$M_{\odot}$ yr$^{-1}$\\
\hline
S$_{Band~7}$$^d$&7.55$\pm$0.15&1.70$\pm$0.19&1.58$\pm$0.18&mJy\\
$L_{TIR}$ (Band 7)$^e$&8.8&1.66&1.65&$10^{12}$~$L_{\odot}$ (8-1000 $\mu$m)\\
SFR (Band 7)&1310$\pm$263&247$\pm$56&245$\pm56$&$M_{\odot}$ yr$^{-1}$\\
\hline
S$_{Band~9}$ (host)&$34.57\pm1.37$&&$7.66\pm1.53$&mJy\\
S$_{Band~9}$ (point source)&$<2.8^c$&&$1.01\pm0.41$&mJy\\
$\%$ point source$^c$&$<8$&&11.6\\
$r_{eff, host}$ (Band 9)&$1.02\pm0.06$&&$0.81\pm0.39$&kpc\\
$\nu_{Spergel}$&0.3 (fixed)&&$-0.71\pm0.06$&\\
\hline
\end{tabular}
\begin{tablenotes}
    \item \footnotesize$^a$ CO fits are based on a free Spergel index.
    \item \footnotesize$^b$ Continuum fits are based on the best-fit index to the CO emission with the exception of J2317-1033 which did not converge. In parenthesis, we report the values based on a fixed Spergel index set to the mean value (-0.3) of the sample in \citet{Tan2024b} and also using a free index.
    \item \footnotesize$^c$ Upper limits are evaluated at the 3$\sigma$ error on the point source contribution.
    \item \footnotesize$^d$ Updated band 7 fluxes based on model fits in $uv$-space
    \item \footnotesize$^e$ Host galaxy; Uncertainties are $\sim20\%$.
\end{tablenotes}
\end{threeparttable}
\end{table*}

\bsp
\label{lastpage}
\end{document}